\documentclass{iopart}
\usepackage{graphicx}
\usepackage[T1]{fontenc}  
\begin{document}

\title[Interlayer coupling in EuS/SrS ...]{Interlayer coupling in 
EuS/SrS, EuS/PbSe and EuS/PbTe magnetic semiconductor superlatices}

\author{H K\k{e}pa$^{1,2}$, C F Majkrzak$^3$, A Sipatov$^4$, A G Fedorov$^5$,\\
T A Samburskaya$^4$ and T M Giebultowicz$^2$}
\address{$^1$ Institute of Experimental Physics, 69 Ho\.za Str., 00-681
Warsaw, Poland}
\address{$^2$ Physics Department, Oregon State University, 301 Weniger Hall,
Corvallis, OR 97331, USA}
\address{$^3$ NIST Center for Neutron Research, National Institute of Standards
and Technology, Gaithersburg, MD 20899, USA}
\address{$^4$ National Technical University KPI,
21 Frunze Str., 61002 Kharkov, Ukraine}
\address{$^5$ Institute for Scintillation Materials NASU, 60 Lenin Ave.,
61001 Kharkov, Ukraine}
\ead{Henryk.Kepa@fuw.edu.pl}
\begin{abstract}
Neutron reflectivity studies of EuS/SrS, EuS/PbSe, and EuS/PbTe 
all-semiconductor superlattices were carried out in search for
exchange interlayer coupling.
A relatively weak antiferromagnetic coupling was found in
EuS/SrS and in EuS/PbSe systems but no interlayer coupling was
detected in EuS/PbTe superlattices.
In EuS/SrS, where the SrS spacer is an insulator ($E_g\approx 4$ eV), 
a very weak and short range interlayer coupling is in agreement with the
earlier theoretical predictions that the interlayer coupling strength 
in EuS-based magnetic semiconductor superlattices depends strongly
on the energy gap of the nonmagnetic layer and should decrease
with an increase of the energy gap of the spacer material.
A weak coupling in EuS/PbSe and no coupling in EuS/PbTe, where both PbSe 
and PbTe are narrow-gap semiconductors ($E_g\approx 0.3$ eV), 
is in disagreement not only with the theoretical expectations but also 
in a stark contrast with earlier results for another narrow-gap spacer 
system -- EuS/PbS, where pronounced antiferromagnetic coupling persists 
even in samples with PbS layer thickness as large as 200 \AA.  
A possible influence of the increasing lattice mismatch between EuS and 
the spacer materials (0.5\%, 0.8\%, 2.5\%, and 8.2\% for PbS, SrS, PbSe, and
PbTe, respectively) 
on the magnetic in-plane ordering of EuS layer and, consequently, on the 
interlayer coupling was investigated by polarized neutron reflectometry in the
case of EuS/PbTe.  
\end{abstract}

\maketitle

\twocolumn

\section{Introduction}
\label{intro}
EuS is a well-known ferromagnetic
semiconductor. It has been extensively studied since the 1960s, so that its
magnetic and electronic properties have been  well characterized
\cite{wachter}.

Because of its low Curie temperature, EuS ($T_c=16.5$ K) 
is certainly not a good candidate for practical spintronics
applications. However, since its magnetic and electronic properties are
well understood, it may serve as a good ``prototype'' for investigating
the fundamental physical mechanisms  underlying phenomena that are 
highly interesting  in the context of spintronics studies. 
Interlayer magnetic coupling in all-semiconductor superlattices is one
such topic.

This work is a continuation of earlier studies of interlayer magnetic coupling
in EuS/PbS and EuS/YbSe semiconductor superlattices
\cite{kepa_epl,kepa-jac} in which
a pronounced antiferromagnetic (AFM) interlayer coupling (IC) has been
found. All the present and the previous neutron reflectometry experiments
were carried out at the NIST Center for Neutron Research, Gaithersburg, USA
on the NG-1 neutron reflectometer.

\section{EuS/SrS superlattices}
The origin of IEC phenomena in all-metallic
multilayers is now quite well understood. As shown by a number of theoretical 
studies backed by experimental results (see,  e.g., \cite{bruno}),
the interaction
between two metallic FM blocks across a nonmagnetic metallic spacer is
maintained by conduction electrons. 
However, this mode of interaction cannot give rise to any observable 
interlayer coupling effects in superlattices composed of semiconduting materials
in which  the concentration of mobile carriers is orders of magnitude
lower than in metals.

In order to explain 
the origin of the  pronounced interlayer coupling  seen in all-semiconductor 
SL systems such as EuS/PbS, EuS/YBSe, and EuTe/PbTe,
Blinowski and Kacman (B\&K) proposed a model in which the exchange 
interactions are conveyed across the semiconductor spacers 
by {\it valence band electrons}~\cite{blin2}. The model does 
not assume any particular interaction mechanism, but attributes the
interlayer coupling to the sensitivity of the superlattice electronic energies
to the magnetic order in the layers --  i.e., it accounts globally for
the spin-dependent band structure effects.
According to the Blinowski and Kacman model, the strength of the coupling
between the EuS layers decreases exponentially with the spacer layer
thickness. 
The model also predicts that that exponential decrease is faster for the 
systems where the energy gap of the spacer material is larger \cite{sank1}.
In other words, 
weaker and  shorter range interactions should be seen in EuS/YbSe 
($E_g=1.6$ eV for YbSe) multilayers than in EuS/PbS ($E_g=0.2$ eV 
for PbS) superlattices. This has been demonstrated experimentally by neutron 
reflectometry experiments performed in applied magnetic fields \cite{kepa-jac}.

In order to further test the predictions of the B\&K theory, we have
carried out neutron reflectometry studies of EuS/SrS superlattices.
\begin{figure}
\centering
\includegraphics[width=6cm]{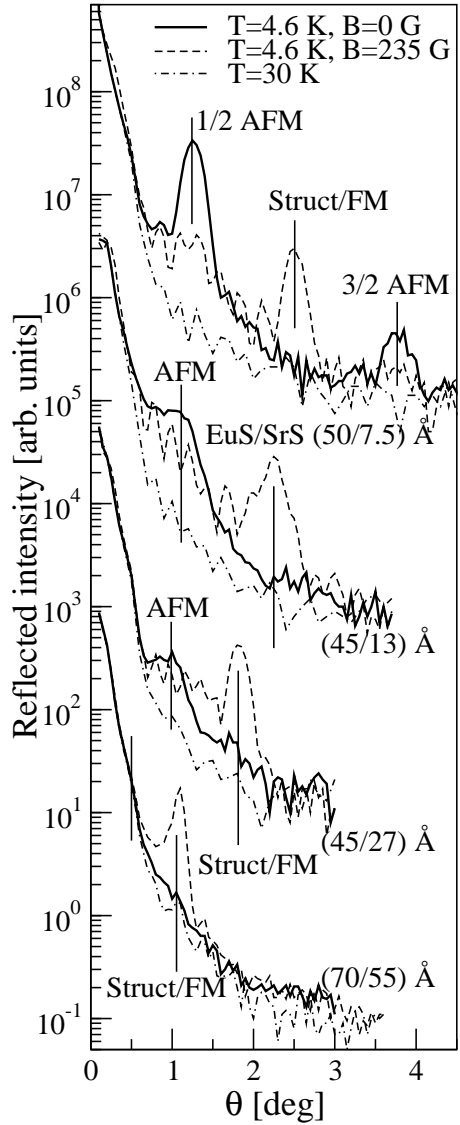}
\caption{Unpolarized neutron reflectivity profiles for four EuS/SrS 
superlattices with increasing thickness of the SrS layer. Data taken
below and above the bulk-EuS Curie point (16.6 K), solid and dot-dashed lines,
respectively, and in zero and saturating (dashed line), 235 G, magnetic fields.}
\label{srs-4scans}
\end{figure}
All three spacer materials, PbSe, YbSe and SrS, have the same structure
(NaCl type) as EuS, and are nearly perfectly lattice-matched
with EuS. The strain effects in the EuS/SrS system thus remain essentially 
the same as in EuS/PbS and EuS/YbSe superlattices.  
PbS is a narrow-gap semiconductor, YbSe is a wide-gap semiconductor,
and SrS ($E_g=4.2$ eV) is usually cosidered to be an insulator. 
Therefore, any changes in the strength and 
the range of IC in these three systems can be attributed, with a high degree 
of certainty, only to the differences
in electronic band structure of the spacer material.

The results of neutron reflectivity measurements performed on four EuS/SrS 
superlattices 
with different structural parameters are
presented in the Figure~\ref{srs-4scans}. The dot-dashed lines represent the
reflectivity profile measured at 30 K, which assures that EuS is still in its 
paramagnetic phase. The solid lines show the reflectivity measured at 4.6 K 
(well below the Curie temperature of $T_{\rm C}= 16.5$ K) and in zero external 
magnetic field. In the case of the 
thinnest (7.5 {\AA}) SrS spacer, two half-order ($\frac{1}{2}$ and 
$\frac{3}{2}$) superlattice reflections prove the antiparallel alignment of
adjacent EuS layer magnetizations.
 These AFM maxima disappear in sufficiently strong (saturating) applied 
magnetic field where the magnetization
vectors in all EuS layers are aligned parallel to the field. 
In such a situatuation, the magnetic Bragg maxima occur at the same 
positions as the structural maxima -- in practice it means that the 
structural Bragg peaks should increase in intensity due to the additional 
ferromagnetic (FM) contribution. It should be noted, however,  that EuS/SrS 
is a special case of a superlattice system -- namely, due to the
very small nuclear scattering contrast between EuS and SrS layers
(nuclear scattering lengths of Eu and Sr are $7.22-1.26i$ fm and $7.02$ fm, 
respectively) the structural SL Bragg peaks are  extremely weak and they 
are not visible against the background in neutron reflectivity spectra 
(they can be  seen only in X-ray reflectivity spectra, and from such 
measurements one can accurately determine their positions). Magnetic and
structural properties of EuS–SrS semiconductor multilayers were studied earlier
by SQUID, magneto-optical Kerr effect magnetometry, and by 
X-ray diffraction methods \cite{szot}. 
  
On the other hand, due to the 
very large magnetic moment of the Eu$^{2+}$ cation (7 $\mu_{\rm B}$), 
the magnetic scattering contrast between EuS and 
nonmagnetic SrS is quite strong.  Concsequently, the FM peaks 
measured at saturating field are clearly visible and well developed,
which attests to the good structural quality of the samples.

The AFM SL peak becomes less and less  pronounced as the SrS spacer layer
grows thicker (to 13 and 27 {\AA}), reflecting the decrease
of the AFM interlayer 
coupling strength with increasing spacer thickness. For the sample with
the SrS layer 55 {\AA} thick there is no longer any visible trace of the 
AFM interlayer ordering.

\section{Interlayer coupling in EuS/PbSe and EuS/PbTe superlattices}
\begin{figure}
\centering
\includegraphics[width=6cm]{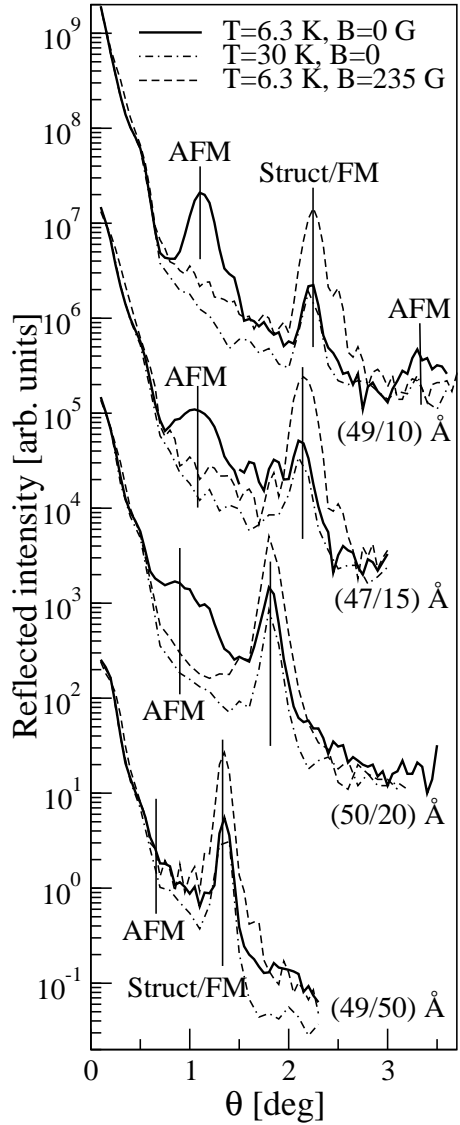}
\caption{Unpolarized neutron reflectivity profiles for several EuS/PbSe 
superlattices with increasing thickness of the PbSe spacer. Data taken
below (solid line) and above (dot-dashed line) the EuS Curie point (16.6 K) 
and in zero and saturating, 235 G,
(dashed line) magnetic fields.}
\label{pbse4scans}
\end{figure}

A similar experimental procedure, to that described above, was applied to the 
set of EuS/PbSe superlattices. The results obtained from those measurements 
are summarized in Figure~\ref{pbse4scans}. 
In contrast to the EuS/SrS data, the spectra from EuS/PbSe specimens exhibit 
well-developed structural SL Bragg peaks. This comes from the fact that 
the scattering length values for both constituent elements of the spacer layers
are larger than the scattering length values of Eu and S
($b_{coh}^{Pb}=9.405$ fm and and $b_{coh}^{Se}=7.970$ fm, compared to   
$b_{coh}^{Eu}=7.22-1.26i$ and  $b_{coh}^{S}=2.847$ fm),
giving rise to a much larger scattering length density contrast than in
the case of EuS/SrS system.

The presence of clearly visible maxima at AFM positions is clear evidence
for the 
existence of significant AFM interlayer coupling in the EuS/PbSe system.
The dependence of the AFM peak intensity on the spacer thickness 
closely resembles the situation seen in the EuS/SrS SL system. However,
one would expect much stronger IC effects in EuS/PbSe, considering that
the PbSe spacer is a narrow gap semiconductor with an $E_g$ value very close 
to that in PbS -- and in the EuS/PbS system  much stronger and
longer range IC was in fact observed \cite{kepa-jac}.

\begin{figure}
\centering
\includegraphics[width=6cm]{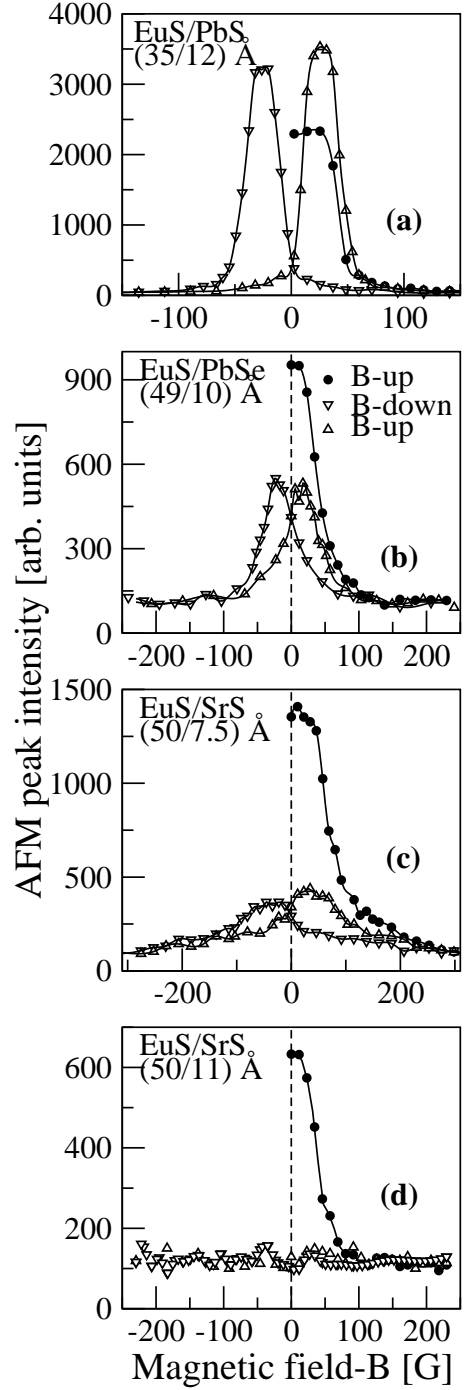}
\caption{The intensity of the AFM SL Bragg peak vs. magnetic field
(``AFM-hysteresis loops'') for (a)-EuS/PbS, (b)-EuS/PbSe, 
and (c-d)-EuS/SrS SL's.}
\label{hyst}
\end{figure}

As it was decribed earlier \cite{kepa-jac,sank2}, 
in order to evaluate quantitatively the
IC strength the AFM peak intensity was measured as a 
function of applied magnetic field.
The results of such experiments performed on 
EuS/PbS, EuS/PbSe, and EuS/SrS samples
are shown in Fig.~\ref{hyst}.
The measurement procedure is the following: the sample is first cooled down to 
the desired temperature in zero magnetic field, and then $T$ is kept constant 
throughout the entire experiment. The intensity of the AFM peak is recorded.
Then a magnetic field $B$ is applied and increased in constant increments, and
the AFM peak intensity is recorded for each $B$ value.
The procedure is continued until $B$ reaches the saturation value (i.e., where
the AFM peak completely disappers).
The field is then decreased in the same manner back to zero. Next, the field
direction is reversed, again increased to the saturation value, and decreased 
back to zero -- then again reversed, and one more up-down cycle is performed. 
This allows us to measure the 'hysteresis
loop' of the antiferromagnetically coupled superlattice.

All samples we have investigated so far (with the exception of one  EuS/PbS 
(30/4.5) {\AA} specimen which showed essentially no hysteresis effects) 
remained in the {\em ferromagnetic} configuration
after the first cycle of the field. Further application of the magnetic 
field in the opposite direction leads to a partial restoration of the 
intensity of the AFM peak. From the latter it can be inferred that the 
antiferromagnetic interlayer configuration can be restored only in a 
fraction of the sample.
How large this fraction is depends on the nonmagnetic
spacer type (PbS, YbSe, PbSe or SrS) and its thickness. Thus, for a 
sufficiently thin PbS spacer
(12 \AA, see Figure~\ref{hyst}(a)) one can observe a full restoration
of the initial AFM state (in this particular example the fraction of the
AFM coupled sample is even higher than it was in the initial 
zero-field-cooled state).
In the case of PbSe, another narrow-gap semiconductor spacer, the degree of 
restoration is 
lower than for a PbS layer of approximately the same thickness 
(see Fig.~\ref{hyst}(b)).
This suggests that the AFM coupling strength in the EuS/PbS
samples is considerably stronger than in the EuS/PbSe samples 
with the same spacer thickness.
In the case of SrS, which is actually an insulator, the degree of recovery of
AFM intensity is far 
lower than for PbS and lower than for PbSe spacers (see Figure~\ref{hyst}(c))
despite the fact that the
thickness of SrS layer is considerably smaller, only 7.5 {\AA}. 
The weakness of IC strength in EuS/SrS is most dramatically visualized 
in Figure~\ref{hyst}(d) for 
EuS/SrS sample with an 11 {\AA} spacer, where there is no recovery of the AFM
peak. 

Thus, in summary, for superlattices with PbS, YbSe and SrS spacers, 
which  are all very well lattice-matched with EuS, interlayer interactions 
systematically decrease as the energy-gap of the spacer material increases.
This finding fully corroborates the theoretical predictions of 
Blinowski-Kacman tight-binding calculations \cite{blin2,sank1}.

\begin{figure}
\centering
\includegraphics[width=6cm]{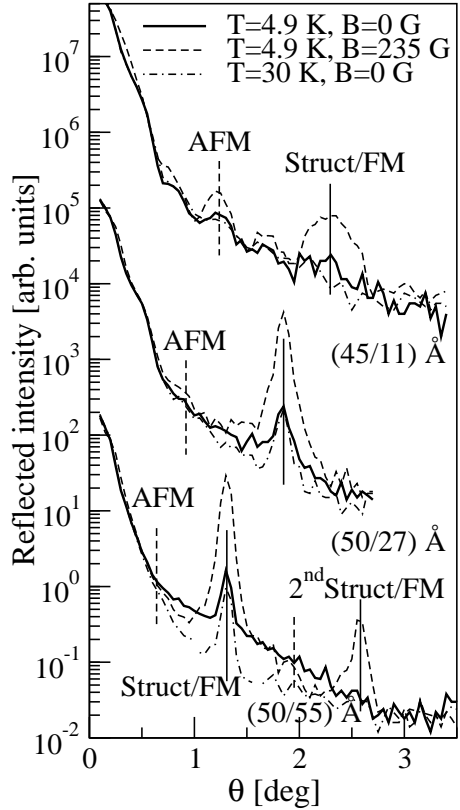}
\caption{Unpolarized neutron reflectivity profiles taken below and above
EuS Curie temperature ($T_{\rm C}=16.6$ K) in zero and saturating (235 G)
magnetic field for three EuS/PbTe samples with different thicknesses
of the nonmagnetic spacer PbTe.}
\label{pbte3scans}
\end{figure}

The results of unpolarized neutron beam reflectivity measurements on the
EuS/PbTe system are presented in Figure~\ref{pbte3scans}. 
As can be seen from these plots,
no AFM interlayer ordering was detected even for
the thinnest PbTe spacer. This appears to be in a stark contrast with
previous results from  narrow-gap spacer systems, EuS/PbS and EuS/PbSe, where
pronounced IC was detected. 
The presence of chemical structural SL Bragg
peaks and large FM contributions in applied magnetic field attest to
the good quality of the SL specimens. The bulk unstrained lattice parameters 
of EuS, PbS, PbSe, and PbTe are, respectively, 
 5.968 {\AA}, 5.936 {\AA}, 6.12 {\AA}, and 6.460 {\AA},
so the lattice misfits between EuS and the three spacer 
materials are about 0.5 \%, 2.5\%, 8\%, respectively.
The apparently weaker IC in EuS/PbSe compared to EuS/PbS, and the fact that no
interlayer coupling is seen in EuS/PbTe, clearly suggest that there 
exists a correlation -- the larger the lattice misfit strain, the 
lower the degree of AFM interlayer correlations in the system. 
This observation will be addressed in closer detail in the next section.

\section{Misfit strain in EuS/PbSe and EuS/PbTe superlattices}
In the process of epitaxial deposition of one monochalcogenide on the top of
a layer of another monochalcogenide, the deposited layer's in-plane
interatomic distances initially coincide with those of the substrate. In this
so called
pseudomorphic stage, the layer with the larger lattice parameter is compressed
and the layer with the smaller one is expanded. The value of the stress depends
on the misfit parameter $f=2(a_1-a_2)/(a_1+a_2)$ between lattice constants
$a_1$ and $a_2$ of the two materials. When the thickness of the deposited layer
exceeds some critical magnitude $d_c$, the accumulated elastic energy of the
layers is partially released through the formation of misfit dislocations. 
The higher the misfit $f$, the smaller the critical thickness $d_c$,
(see e.g., \cite{markov,venables,jain}).

\begin{figure}
\includegraphics[width=7cm]{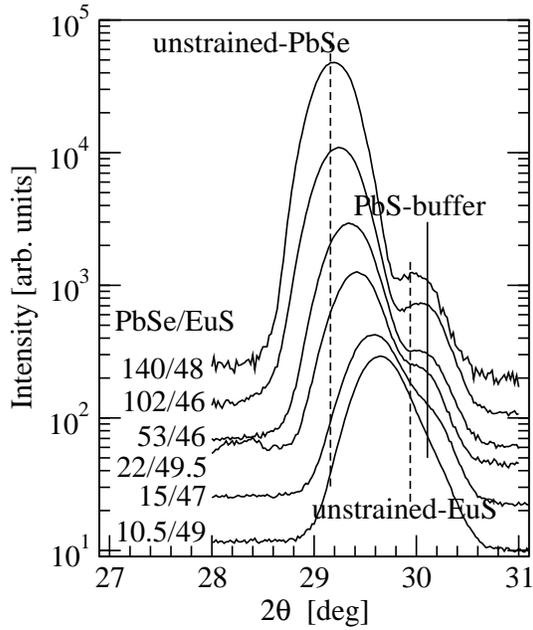}
\caption{X-ray (200) reflection, measured in transmission geometry, from a
number of a pseudomorphic EuS/PbSe superlattices with approximately constant
EuS and increasing PbSe layer thicknesses. Apart from the PbS buffer peak, there
is only one maximum, which confirms the coincidence of the in-plane lattice
parameters of EuS and PbSe. The value of the common in-plane lattice parameter
shifts towards the unstrained (bulk) PbSe lattice constant as the thickness
of the PbSe layer increases.}
\label{x-pbse}
\end{figure}

In the case of EuS/PbSe systems the crical thickness is about 150 \AA{}; thus
all superlattices with layer thicknesses below this value grow
pseudomorphically with the common in-plane lattice parameter. 
The results of X-ray diffraction experiments carried out on a number of
EuS/PbSe SL's with
approximately constant EuS layer thickness, equal to about 50 \AA{}, and
PbSe spacers in the range from 10 to 140 \AA{} are shown  in Figure
\ref{x-pbse}. The (200) reflection obtained in transmission 
geometry yields the value of the in-plane, common for the both EuS and
PbSe layer,
lattice parameter. The diffraction maximum shifts towards the bulk-PbSe
(200) peak position as the thickness of the PbSe layer increases, clearly
showing that EuS layer is under increasing tensile, in-plane, strain from the
increasingly thicker PbSe spacers. 

It is well known that any strain deformation of a magnetic 
multilayer structure significantly modifies its magnetic properties 
(see e.g. \cite{stachow,story2}).
The influence of uniaxial and hydrostatic strain on 
interlayer coupling in EuS-based SL's has been theoretically investigated
in \cite{sank1}. Although the case of in-plane uniaxial strain has not been
considered in that work, the results of model calculations performed for
other distortion types lead to the conclusion that any
deformation which increases the distances between the magnetic ions in the 
SL structure results in a reduction of IC.
Thus, the faster decay of IC strength vs nonmagnetic spacer thickness in
EuS/PbSe compared to EuS/PbS can be attributed to the tensile deformation
of the EuS layers in the EuS/PbSe system. The fact that the deformation
increases with growing PbSe spacer thickness further enhances
the rate at which IC deteriorates.
In EuS/PbS SL's the lattice misfit
is much smaller, the deformations are negligible and the decrease of the IC
vs PbS layer thickness depends on the spacer thickness only.

\begin{figure}
\centering
\includegraphics[width=5cm]{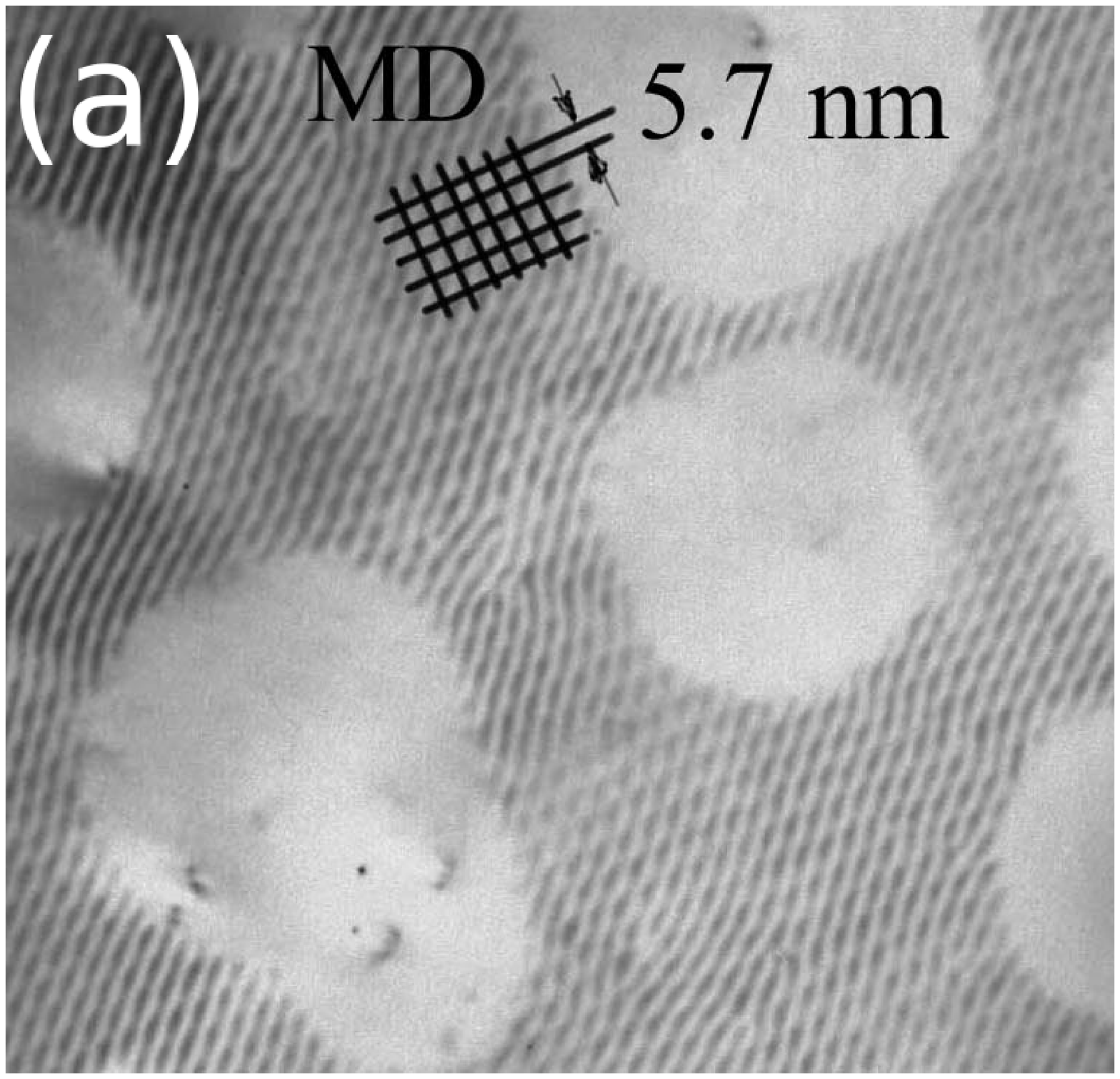}
\includegraphics[width=5cm]{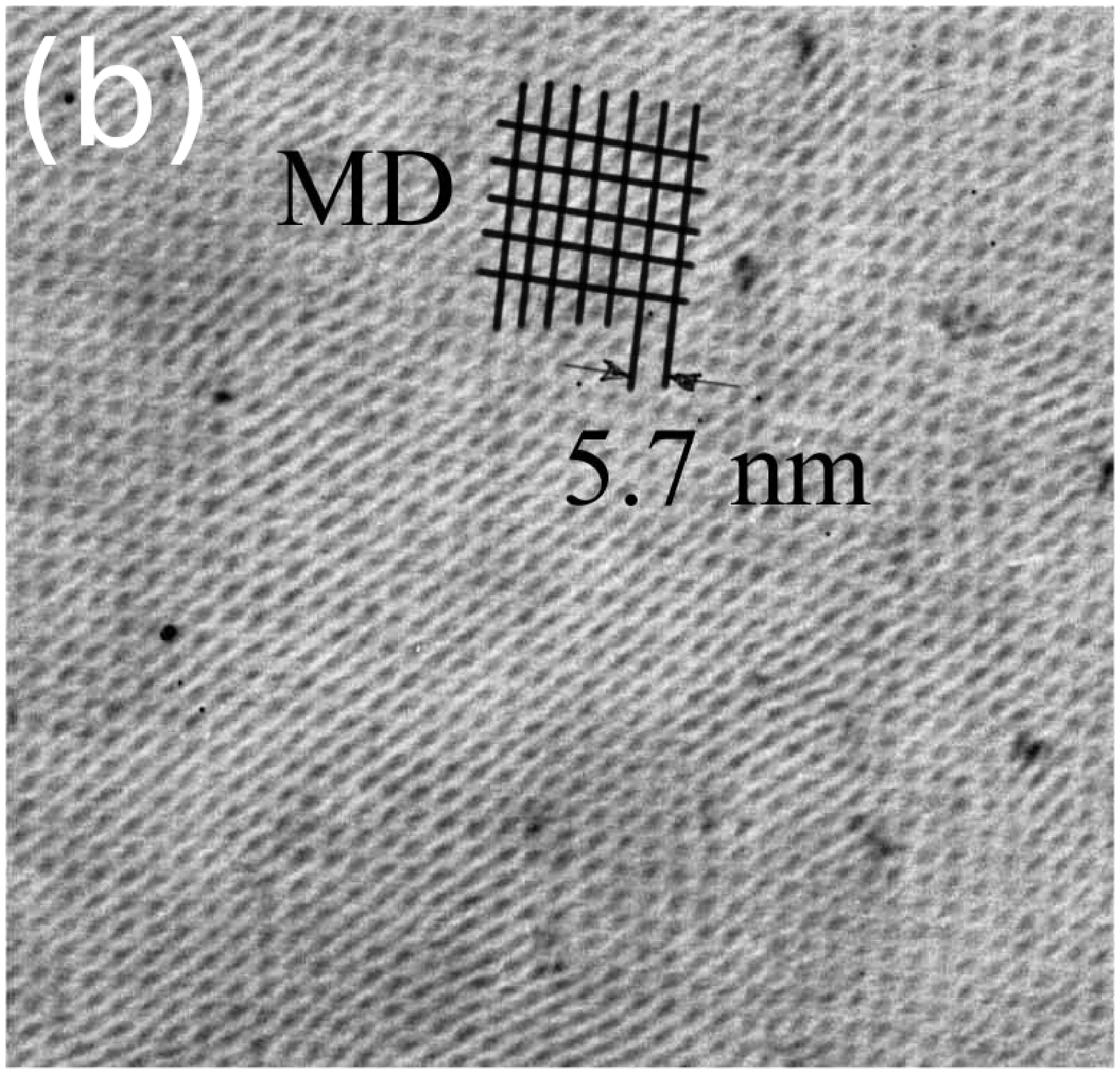}
\includegraphics[width=5cm]{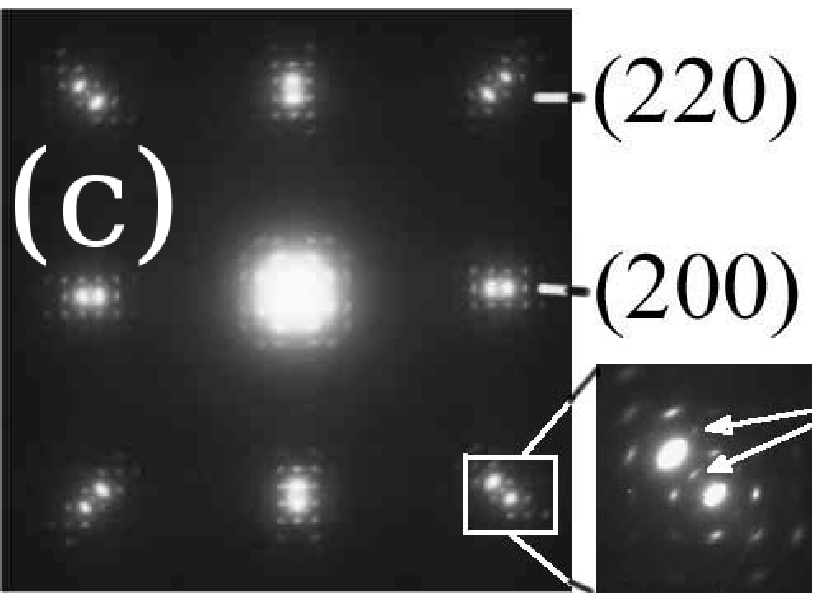}
\caption{(a) Electron microscopy image of the Stransky-Krastanov
 mode of growth (island-like) of a thin PbTe layer on EuS;
(b) electron microscopy image of the EuS-PbTe bilayer demonstrating
the long-range square misfit dislocation grid for the thicker, continuous,
layers; (c) electron diffraction showing
the satellites, accompanying the fundamental (200) and (220) reflections
(see the arrows in the inset), arising from the periodic arrangement of
dislocation lines at the interface.}
\label{dislocgr}
\end{figure}

A completely different situation takes place in the EuS/PbTe superlattices.
Here, due to the very large misfit, the critical thickness is so small that
only one monolayer of the material (EuS or PbTe) can be grown pseudomorphically
on the other. Thus, in the case of EuS/PbTe SL's one never can obtain a
pseudomorphic system -- instead, after the first monolayer is deposited, the 
growth proceeds in the Stransky-Krastanov mode.
The deposition process produces not a continuous layer, but separate islands.
In each such island
relatively short-range edge misfit dislocation grids are formed,  as shown in 
Figure~\ref{dislocgr}(a). 

As the thickness of the deposited layer increases, the islands coalesce. A 
continuous layer is formed and the misfit dislocations order into a long range
periodic grid at the interface between the two layers. 
The periodicity
of the grid is inversly proportional to the misfit parameter $f$ and
equals 57 {\AA} for EuS/PbTe. 
An example of such a grid formed in a EuS/PbTe 
bilayer can be seen in in the electron microscope image presented in 
Figure~\ref{dislocgr}(b). And in Figure~\ref{dislocgr}(c) 
an electron diffraction image from the same bilayer is presented,
showing a number of superstructure reflections corresponding to the 
periodicity of the dislocation grid.

The dislocation grids release much of the misfit strain in the multilayer.
The remaining unrelaxed strain is now sinusoidally modulated in two ortogonal
directions in the plane of the interface. The distribution of the
elastic deformation of the EuS lattice in the interface plane, is illustrated
in Refs.~\cite{fogel2002,mikhailov2}, (in Figures 10 and 5, respectively).  

One can expect that the presence of such dislocation grids in the EuS
layers may give rise to highly interesting magnetic phenomena. 
The sinusoidally modulated strain exerts alternating  
compresive and tensile pressure on the magnetic material near the interfaces. 
Further away from the interface
regions, in the middle of the layer, the influence of the interfacial strain
may be greatly reduced (see Figure 5 in~\cite{mikhailov2}) leaving that 
portion of the EuS layer essentially
unchanged. The actual strain distribution depends on the lattice misfit and on 
the thicknesses of both, magnetic and nonmagnetic, 
layers~\cite{mikhailov2,fedorenko}.
Neutron diffraction under sufficiently high
hydrostatic pressure has shown that the Curie temperature of EuS
is very sensitive to the applied pressure. A pressure of 20 GPa shifts the
EuS $T_{\rm C}$ from 16.6 K to about 150 K~\cite{goncha}.
The peak value of the interplanar distance modulation due to 8\% 
lattice misfit between EuS and PbTe in a multilayer system is equivalent to 
an applied  uniaxial pressure of about 25 GPa.
Consequently, the near-interface areas of the EuS layer that experience
compresive strain should stay magnetically ordered at temperatures
considerably higher than the Curie point of an unstrained EuS layer.
By the same token, the-near interface portions of the EuS
layer which are under tensile strain should remain magnetically disordered
at temperatures well below the $T_{\rm C}$ of unstrained EuS. 

\section{Polarized neutron reflectivity investigation of EuS/PbTe}
\begin{figure}
\centering
\includegraphics[width=6cm]{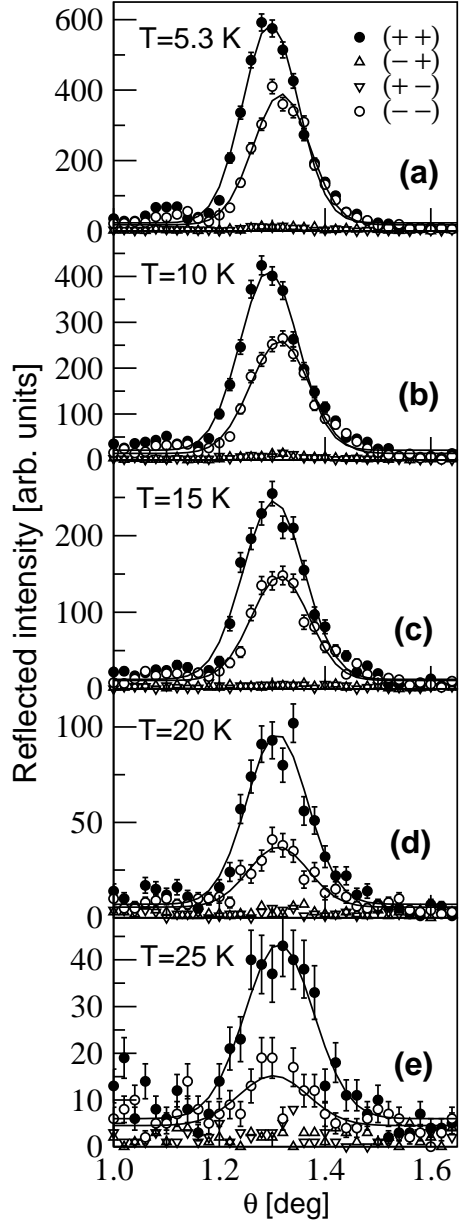}
\caption{(a)-(e) -- Polarization analysis of the first SL Bragg peak 
from a EuS/PbTe (50/50) \AA{} superlattice carried out at 0.5 T external 
magnetic field at different temperatures. All four non-spin-flip and spin-flip
 scattering processes $(+ +)$, $(- +)$, $(+ -)$, and $(- -)$ are presented.}
\label{pol_pbte}
\end{figure}
Our attempts to directly demonstrate the magnetization modulation by magnetic 
neutron diffraction measurements have so far been
unsuccessful. A possible reason may be the relatively low 
volume of the stress modulated EuS. A significant modulation may only be
confined to relatively small portions of the EuS layers in the vicinity of
the PbTe-EuS interfaces -- which results in a very low diffracted intensity.
Below $T_{\rm C}$ larger, unmodulated part of the layer plays
a dominant role, masking
the much weaker signal produced by the smaller modulated part on the diffracted 
intensity. Above $T_{\rm C}$ the main, unmodulated, part of the layer becomes
magnetically disordered, and only the most compressed regions still remain
magnetically ordered, forming a regular array of ``nanomagnets''. It does not 
appears likely that such "nanomagnets" can interact with each other across
the unmagnetized regions separating them. 
Therefore, in order to observe  
coherent magnetic diffraction from  the array one should apply a sufficiently 
strong magnetic field that would allign all the magntization vectors in
individual ``nanomagnets''.

However, even in such circumstances the detection of a magnetic diffraction
signal
from the array poses a substantial challenge because this maximum occurs at 
the same position as the overwhelmingly stronger nuclear Bragg peak from
the SL structure. Fortunately, the technique of polarized neutron diffraction
offers the possibility to extract small magnetic
signals superimposed on strong nuclear diffraction maxima. In brief,
this method is based
on the fact that for non-spin-flip scattering of a polarized neutron beam from
a magnetized ferromagnet, the magnetic and the nuclear scattering components
interfere -- either constructively or destructively, depending on whether the
neutron spins are parallel, or antiparallel to the sample magnetization vector.
The difference between the intensities observed in these two scattering modes --
which are conventionally denoted as $(++)$ and $(--)$ is proportional to the 
sample magnetization. 

Figure~\ref{pol_pbte} presents the results of polarization analysis applied to
the 1$^{st}$ order SL Bragg peak from the EuS/PbTe (50/50)~{\AA} superlattice.
The sample was placed in an applied magnetic field of 0.5 Tesla. All four
scattering processes, the two non-spin-flip $(++)$ and $(--)$, and the two 
spin-flip $(-+)$ and $(+-)$ have been measured. (The NIST NG-1 cold neutron
reflectometer
in polarized mode of operation was used.) As can be seen from the 
Figure~\ref{pol_pbte}, the very low and flat spin-flip, $(-+)$ and $(+-)$, 
intensities prove that the sample is in a saturated state, with no 
transverse components of the magnetization. 
\begin{figure}
\centering
\includegraphics[width=6cm]{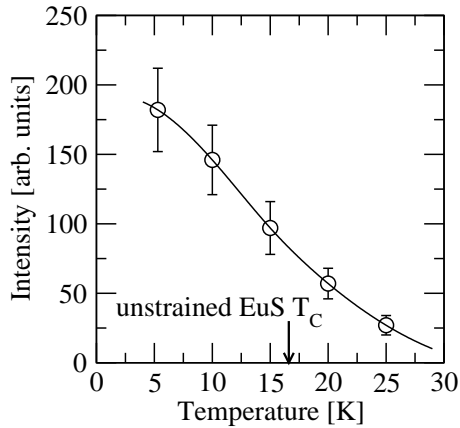}
\caption{The difference between the $(++)$ and $(--)$ non-spin-flip
intensities vs. temperature. The solid line is a guide for the eye only.}
\label{spinsplit}
\end{figure}

The difference between the non-spin-flip, $(++)$ and $(--)$, intensities is 
proportional to the sample magnetization~\cite{moon}. 
In Figure~\ref{spinsplit} the difference between $(++)$ and $(--)$ 
intensities is plotted vs. temperature.
As can be seen from
Figure~\ref{pol_pbte}(d)--(e) and Figure~\ref{spinsplit}, the 
magnetic contribution to the 
scattering persists well above the Curie point of the unstrained EuS (16.6 K).

\section{Summary}
In this paper the trends in interlayer coupling for two EuS-based
families of all-semiconductor superlattices were investigated.

In one group
of the studied SL's, EuS/PbS, EuS/YbSe, and EuS/SrS, the nonmagnetic spacer
materials are very well lattice matched (the misfit $f\approx 0.5\%$) to EuS.
Due to the very small $f$, and consequently large $d_c$, the layers in all 
these superlattices grow pseudomorphically and are negligibly stretched/expanded
and have no dislocation grids formed on the interfaces. As a result, the EuS
layers
are magnetized uniformly, without modulation in the interface vicinity.
The energy-gap of these spacers systematically increases from about 0.3 eV for
narrow-gap PbS, through ~1.6 eV for wide-gap YbSe, to about 4 eV for 
insulating SrS. The experimentally observed strength of the interlayer 
coupling in this
series of SL's decreases monotonically as the energy gap of the spacer 
increases. These findings strongly support the prediction of
the tight-binding calculations by Blinowski and Kacman~\cite{blin2,sank1}.

In the other group of studied SL's, EuS/PbS, EuS/PbSe, and Eus/PbTe, all the
spacers are narrow gap semiconductors with an energy gap of the order of
0.2 - 0.3 eV and have lattice misfits to EuS which monotonically increase from 
about 0.5\% for PbS through 2.5\% for PbSe to about 8\% for PbTe. 
Although, the EuS/PbS and EuS/PbSe systems are both pseudomorphic, the 
considerably higher tensile strain of EuS in EuS/PbSe SL's leads to a noticible
decrease of the IC strength in the latter system as compared with EuS/PbS SL's.
In EuS/PbTe no AFM IC was observed, most likely due to the drastic 
modifications of the EuS magnetic properties around the interfaces. 
In addition, Stransky-Krastanov, island-like, growth of the layers contribute 
to the increased
roughness of the interfaces and, very likely, discontinuities or  
pin-holes in the thinner PbTe spacers. 
The latter may be responsible for the weak ferromagnetic interlayer coupling,
due to the direct contact between adjacent EuS layers,
seen in some EuS/PbTe SL samples.

An enhanced EuS $T_{\rm C}$ in EuS/PbTe superlattices has been
found by neutron polarization analysis performed, on the first-order SL
Bragg peak, in a saturating magnetic field. The origin of this enhancement
was attributed to the existence of highly compressed EuS regions in the vicinity
of EuS-PbTe interfaces due to the misfit dislocation grid in EuS/PbTe system.

\ack
The work was supported by NSF DMR-0204105, NSF DMR-0509478
and CRDF UKP2-2896-KV-07 grants.

\section*{References}

\begin{thebibliography}{10}

\bibitem{wachter}
P.~Wachter,
\newblock Europium chalcogenides: {EuO}, {EuS}, {EuSe} and {EuTe},
\newblock in {\em Handbook on the Physics and Chemistry of Rare Earth}, edited
  by K.~A. Gschneider and L.~R. Eyring, volume~2, pages 507--574, North
  Holland, Amsterdam, 1979.

\bibitem{kepa_epl}
H.~K\k{e}pa et~al.,
\newblock Europhys. Lett. {\bf 56}, 54 (2001).

\bibitem{kepa-jac}
H.~K\k{e}pa, C.~F. Majkrzak, P.~Sankowski, A.~Y. Sipatov, and T.~M.
  Giebultowicz,
\newblock J. Alloys and Compounds {\bf 401}, 238 (2006).

\bibitem{bruno}
P.~Bruno,
\newblock Phys. Rev. B {\bf 52}, 411 (1995).

\bibitem{blin2}
J.~Blinowski and P.~Kacman,
\newblock Phys. Rev. B {\bf 64}, 045302 (2001).

\bibitem{sank1}
P.~Sankowski and P.~Kacman,
\newblock Acta Phys. Polon. A {\bf 103}, 621 (2003).

\bibitem{szot}
M.~Szot et~al.,
\newblock Acta Phys. Polon. A {\bf 112}, 419 (2007).

\bibitem{sank2}
P.~Sankowski et~al.,
\newblock Acta Phys. Polon. A {\bf 105}, 607 (2004).

\bibitem{markov}
I.~V. Markov,
\newblock {\em Crystal Growth for Beginners: Fundamentals of Nucleation,
  Crystal Growth, and Epitaxy},
\newblock World Scientific, Singapore, 2003.

\bibitem{venables}
J.~Venables,
\newblock {\em Introduction to Surface and Thin Film Processes},
\newblock Cambridge University Press, Cambridge, 2000.

\bibitem{jain}
S.~Jain, M.~Willander, and R.~V. Overstraeten, editors,
\newblock {\em Compound Semiconductors Strained Layers and Devices},
\newblock Kluver, Boston, 2000.

\bibitem{stachow}
A.~Stachow-W{\'o}jcik et~al.,
\newblock Phys. Rev. B {\bf 60}, 15220 (1999).

\bibitem{story2}
T.~Story,
\newblock physica status solidi (b) {\bf 236}, 310 (2003).

\bibitem{fogel2002}
N.~Y. Fogel et~al.,
\newblock Phys. Rev. B {\bf 66}, 174513 (2002).

\bibitem{mikhailov2}
I.~F. Mikhailov, B.~A. Savitskii, A.~Y. Sipatov, A.~I. Fedorenko, and L.~P.
  Shpakovskaya,
\newblock Sov. Phys. Solid State {\bf 25}, 668 (1983).

\bibitem{fedorenko}
A.~I. Fedorenko, B.~A. Savitskii, A.~Y. Sipatov, and L.~I. Shpakovskaya,
\newblock Sov. Phys. Crystallogr. {\bf 27}, 569 (1982).

\bibitem{goncha}
I.~N. Goncharenko and I.~Mirebeau,
\newblock Phys. Rev. Lett. {\bf 80}, 1082 (1998).

\bibitem{moon}
R.~M. Moon, T.~Riste, and W.~C. Koehler,
\newblock Phys. Rev. {\bf 181}, 920 (1969).

\end{thebibliography}

\end{document}